\documentclass[prd,twocolumn,showpacs]{revtex4-2}
\pdfoutput=1

\usepackage{graphicx,amsmath,amssymb,amsthm}
\usepackage[hidelinks]{hyperref}
\usepackage{xcolor}
\usepackage{ulem}
\usepackage{thmtools, thm-restate}
\usepackage{epsfig}
\usepackage{epstopdf}
\usepackage{latexsym}
\usepackage{graphicx}
\usepackage{booktabs}
\usepackage{bbm}
\usepackage{color}
\usepackage{physics}
\usepackage{tensor}
\usepackage{verbatim}
\usepackage[caption=false]{subfig}
\usepackage{tikz}
\usepackage{bigints}
\usepackage{ifthen}
\usepackage{soul}
\usetikzlibrary{matrix}
\usetikzlibrary{decorations.markings,calc,shapes,decorations.pathmorphing,arrows.meta}
\usetikzlibrary{patterns}
\usetikzlibrary{positioning}
\usepackage{textpos}
\usepackage{hyperref}
\usepackage{xcolor}
\hypersetup{
colorlinks,
linkcolor={blue!80!black},
citecolor={blue!80!black},
urlcolor={blue!80!black},
linktoc=page
}

\newcommand\beq{\begin{equation}}
\newcommand\eeq{\end{equation}}
\newcommand\be{\begin{equation}}
\newcommand\ee{\end{equation}}

\hypersetup{
colorlinks,
linkcolor={blue!80!black},
citecolor={blue!80!black},
urlcolor={blue!80!black},
linktoc=page
}

\begin{document}

\title{Branes Screening Quarks and Defect Operators}
\preprint{today}
\author{Andreas Karch$^{a}$}
\author{Marcos Riojas$^{a,b}$}
\affiliation{$^a$Weinberg Institute, Department of Physics, University of Texas, Austin, TX 78712, USA}
\affiliation{$^b$Center for the Fundamental Laws of Nature, Harvard University, Cambridge, MA, USA}

\begin{abstract}
{
Here we generalize a well-known computation and uncover a phase-transition, showing that Wilson lines do not necessarily exhibit Coulomb scaling laws in AdS/BCFT at zero temperature. The area difference between a surface that returns to the boundary, and one that plunges into the bulk, determines the potential between two quarks. This classic AdS/CFT calculation is naturally extended to Wilson surfaces associated to general p-form symmetries in boundary conformal field theories (BCFTs) by embedding a Karch-Randall (KR) brane in the geometry. We find (generalized) Coulomb law scaling in subregion size $\Gamma$ is recovered only above the critical angle for the brane, $\theta_{c,p}$. The potential between the two quarks (or defect operators) vanishes precisely when the surface connecting them ceases to exist at $\theta_{c,p}$. This screening effect, where the operators are fully screened below the critical angle, is a phase transition from Coulomb law to perimeter law with the brane angle $\theta_b$ acting as an order parameter. This effect is also explored at finite temperature where we introduce a new regularization procedure to obtain closed-form results. 
}
\end{abstract}

\maketitle


\section{Introduction}
Subcritical Randall-Sundrum (RS) branes \cite{Karch:2000ct,Karch:2000gx} with a cosmological constant, in this context also known as Karch-Randall (KR) branes, are practical bottom-up toy models of holographic duals for conformal field theories with boundaries (BCFTs) \cite{Karch:2000ct,Karch:2000gx,Takayanagi:2011zk,Fujita:2011fp}. Particular emphasis has been placed in recent years on the calculation of entanglement entropies in these settings using the Ryu-Takayanagi (RT) prescription \cite{Ryu:2006bv}; crucially, the co-dimension two extremal surfaces whose area determines entanglement entropy may attach to the KR brane. 

It was recently realized that the brane induces a phase transition at late times between two classes of extremal surfaces when the field theory is effectively at finite temperature. This can be achieved for field theories by studying them in thermal equilibrium or by placing them on non-dynamical black hole backgrounds. At early times the entanglement entropy grows linearly in time; from the holographic perspective, this is realized by a horizon-plunging Hartman-Maldacena (HM) surface \cite{Hartman:2013qma}. The presence of the brane allows a time-independent class of saddles -- known as island RT surfaces -- to end on the brane instead of crossing the horizon. Thus the brane prevents the entanglement entropy from growing forever \cite{Almheiri:2019hni,Almheiri:2019psy,Geng:2020qvw,Chen:2020uac,Geng:2020fxl,Chen:2020hmv}. This behavior of the entanglement entropy has been interpreted as a generalized Page curve, by which we mean a time evolution consistent with unitarity. 

\begin{figure}
    \centering
    \includegraphics[width=\linewidth]{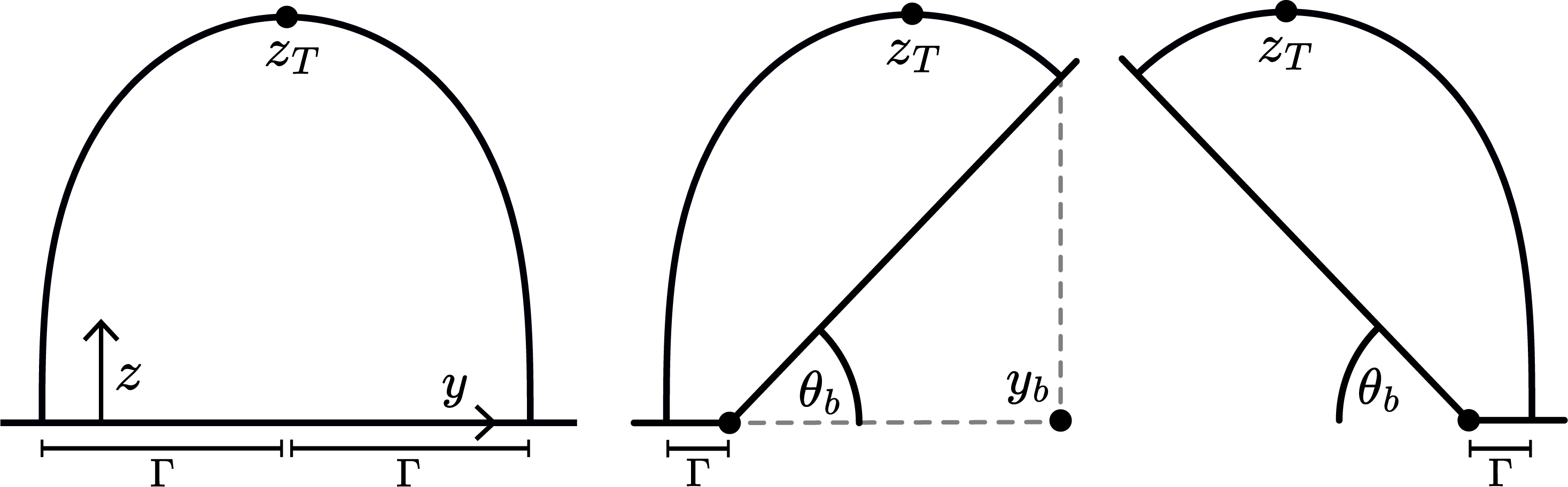}
    \caption{It is well-known that for $p>1$ the $p$-dimensional extremal surface dual to the closed Wilson surface $W(C)$ exhibits Coulomb law as $\Gamma \rightarrow 0$, both for empty AdS and at finite temperature. However, after a KR brane is introduced with angle $\theta_b$, the behavior as $\Gamma\rightarrow0$ depends on whether $\theta_b$ is greater than or less than the critical angle $\theta_c$. Here we depict an ICFT rather than a BCFT to emphasize the close connection with the quark-antiquark potential.}
    \label{fig:Wilson_Line}
\end{figure}

This same phase transition occurs for general $p$-dimensional minimal spatial surfaces in general $d+1$ dimensional locally asymptotically Anti-de Sitter (AdS$_{d+1}$) spacetimes terminating on a KR brane. Instead of a phase transition in entanglement entropy, for an ICFT this can be viewed as a phase transition in the potential between a quark and the boundary or, equivalently, between a quark and its mirror quark, or the analogous potential between a higher dimensional defect and the boundary for $p>2$. For an ICFT this is a transition in the quark-antiquark potential, as depicted in Figure \ref{fig:Wilson_Line}. This is a higher dimensional generalization of the entanglement island formation previously seen in lower dimensions \cite{Almheiri:2019psf,Penington:2019npb,Almheiri:2020cfm} based on the study of quantum entangling surfaces \cite{Engelhardt:2014gca} in theories of gravity coupled to a bath. The brane screens all $p-1$ dimensional defect operators, including Wilson lines $(p=2)$.

These minimal $p-$dimensional extremal objects have standard holographic duals: for $p=1$ (spacelike geodesics) they control two-point functions of large dimension operators; for $p=2$ (Euclidean string worldsheets) they compute spatial Wilson and 't Hooft lines \cite{Maldacena:1998im,Rey:1998ik}; for $p \ge 3$ they realize the corresponding Wilson surface operators. We find that all these observables exhibit the analog of islands and Page curves.

In particular, at zero temperature -- where the island forming phase transition may occur, even though it is not required by unitarity \cite{Chen:2020uac,Geng:2020fxl,Geng:2021mic} -- we find a simple universal pattern. The area of the minimal surface depends only on the surface dimension, $p$, and not on the number of bulk dimensions, $d$. Certain universal aspects are identified that carry through at finite temperature. At zero temperature the Page curve for the entanglement entropy in two-dimensional BCFTs (dual to subcritical RS branes) behaves identically to the 2-point function of heavy operators for BCFTs of dimension $d$, the $d=3$ Page curve gives the behavior of Wilson lines in any dimension, and so on for general $p$. This strongly suggests that BCFTs in general dimension $d$ experience phase transitions at the critical angles $\theta_{c,p}$ \cite{Geng:2020fxl, Chen:2020hmv} for all $p \leq d-1$, provided there is an extended operator associated with this $p$ that can detect the transition.

At zero temperature, we find the expectation value of the $p-1$ dimensional defect operator vanishes at $\theta_{c,p}$ and is independent of subregion size $\Gamma$ below that angle. This is interpreted as the vanishing of the potential between a quark or defect and the boundary. The class of extremal surfaces connecting to the brane ceases to exist when this transition occurs. This is a transition from Coulomb to perimeter law at the critical angle $\theta_{c,p}$ for the brane, with the brane angle acting as an order parameter.  

For finite temperature, the class of extremal surfaces connecting to the brane always exists for spatial Wilson surfaces. For temporal Wilson surfaces this computation gives the potential between quark or defect and boundary and we find it vanishes in the vicinity of the critical angle $\theta_{c,p}$; it  vanishes exactly at $\theta_{c,p}$ only for certain values of $p$. Analogous results are obtained for spatial Wilson surfaces. To show this result, we introduce a new regularization technique  that expresses the area functional in coordinates adapted to the finite angle subtended by the black hole's shadow. Our procedure is simple to use and can sometimes yield closed-form solutions.

The underlying cause of this behavior may have more far-reaching consequences than what will be described here. It was understood long ago that these surfaces exhibit Coulomb scaling laws in AdS/CFT at zero temperature \cite{Maldacena:1998im}. Standard lore says this holds at short-distance even at finite temperature. However, for AdS$_d$ we find that regardless of temperature, generalized Coulomb law scaling is recovered only above the critical angle $\theta_{c,p}$ for the brane, which we extend to general $p$-dimensional surfaces embedded in $d$-dimensional geometry. This occurs because the $p$- dimensional surfaces can shrink to a point only when the brane lies above the critical angle $\theta_{c,p}$. 

Explicit top-down constructions realizing well-known BCFTs with very similar qualitative behavior \cite{DHoker:2007hhe,DHoker:2007zhm,Aharony:2011yc,Assel:2011xz,Gaiotto:2008sa,Bachas:2018zmb} were shown \cite{Uhlemann:2021nhu} to exhibit the critical angle phenomenon; see Appendix \ref{TypeIIBRealization} for details.

\section{The Holographic Calculation}

Here we are interested in minimal area sub-manifolds embedded in the purely spatial geometry of a constant time slice. The region of interest is the near-boundary geometry, i.e. empty AdS$_{d+1}$, so our results are broadly applicable to a wide variety of AdS spacetimes.

This is described by Poincare patch coordinates:

\beq ds^2 = \frac{1}{z^2} (-dt^2 + d\vec{x_T}^2 + d\tilde{y}^2 + d\tilde{z}^2). \eeq

Translations along the defect directions and rotations around the surface insertion in the transverse $x_i$ directions are respected by these surfaces, which satisfy:
\beq t=0, \quad x_i =0, \quad \tilde{y}=\tilde{y}(\tilde{z}) ,\eeq
subject to the \textit{Dirichlet} boundary condition that the surface is inserted at some point $\Gamma$ -- i.e., that $y(0)=\Gamma$.

The action for such a minimal area embedding is:
\begin{equation}
    S = V m \, \int \frac{d\tilde{z}}{\tilde{z}^p} \, \sqrt{1+ (\tilde{y}')^2}   \equiv V_T m A_p(\Gamma) .
    \label{eq:action} 
\end{equation} 
Here $V_T$ is the volume of the surface in the transverse $\vec{x}$ directions.  $A_p(d)$ denotes the actual area per unit transverse volume of the minimal surface anchored at $\tilde{y}=\Gamma$. We will loosely refer to $A_p(d)$ simply as the area throughout. The meaning of $m$ depends on what we are calculating. For an RT surface, $m = \frac{1}{4G}$. 

The action density is completely independent of $d$ and only depends on $p$, the dimension of the submanifold. This is of vital interest in the context of our discussion. As advertised in the introduction, this implies a universality between seemingly unrelated quantities. 

For most of the applications discussed in this work $m$ is a physical tension, and in such cases it is essential for our argument that we are in the ``probe" regime, where $m$ is sufficiently small in Planck units to make the right-hand side of Einstein's equations negligible. These surfaces simply minimize their area to satisfy their classical equations of motion; for an RT surface this is true by definition. Famous examples of probe branes for $p=2$ are F1 and D1 strings of type IIB string theory on AdS$_5$ $\times$ $S^5$ which compute a Wilson and 't Hooft line respectively and only differ in their value of $m$. For $p=3$ the M2 brane of M-theory on AdS$_7$ $\times$ $S^4$ calculates a surface observable in the dual (2,0) theory. The straightforward top-down embedding of these objects is lost once we introduce the KR brane.

As the action \eqref{eq:action} is independent of $\tilde{y}$ we can immediately solve the resulting equations of motion for
$\tilde{y}'(\tilde{z}) = \pm \frac{\tilde{z}^p}{\sqrt{z_T^{2p}- \tilde{z}^{2p}}}$,
where $z_T$ is an integration constant that fixes the position of the turnaround point of the minimal surface. Scale invariance exhibited at zero temperature suggests we rescale by $z_T$, giving dimensionless variables $z = \frac{\tilde{z}}{z_T}$ and $y = \frac{\tilde{y}}{z_T}$:
\begin{equation}
    y'(z) = \pm \frac{z^p}{\sqrt{1-z^{2p}}}.
    \label{eq:EOMzeroT}
\end{equation}
The area density evaluated on the solution is just ${\cal A} = \frac{1}{z^p \sqrt{1-z^{2p}}}$. For later use, $z_b\equiv\tilde{z}_b/z_T$ is where the RT surface connects to the brane at $(y_b, z_b)$ in dimensionless coordinates.

\subsection{The Quark Potential Without Branes}
First we show the standard calculation without branes, extending it to general $p$ when necessary. The UV-infinite area $\int \mathcal{A}$ is regulated with the counter-term $\frac{1}{z^p}$:
\begin{equation}
   \frac{A}{z_T^{1-p} } = \int_0^1 \mathcal{A} \ dz = \frac{\epsilon^{1-p}}{p-1}-\frac{\sqrt{\pi}}{(p-1)} \frac{\Gamma\left(\frac{1}{2}+\frac{1}{2 p}\right)}{\Gamma\left(\frac{1}{2 p}\right)},
   \label{eq:actionintegral}
\end{equation}
where we integrated \eqref{eq:action} up to the turnaround $z_T = 1$ using \eqref{eq:EOMzeroT}. The inter-quark distance is:
\begin{equation}
\Gamma = 2 z_T \left(\int_0^1 y'(z) \ dz \right)= z_T \frac{ \sqrt{\pi} \Gamma \left(\frac{1}{2}+\frac{1}{2p}\right)}{\Gamma\left(\frac{1}{2 p}\right)}.
\label{eq:proportional}
\end{equation}
As indicated in Figure \ref{fig:Wilson_Line}, we must integrate up to $z_T$ on both sides of the turnaround; this gives the factor of 2.
Notice $\Gamma$ is proportional to $z_T$. If we were computing entanglement entropy, this would be the subregion size. The action $A$ is $V E$, where $E$ is the energy of the quark-antiquark pair. The infinite part arises from the quark masses; regulating the action is exactly analogous to subtracting off the HM surface. The finite difference is the quark-antiquark potential, $V_{q \bar{q}}$: 
\begin{equation}
    V_{q \bar{q}} \equiv 2 \Delta A =  - \frac{\sqrt{\pi}}{(p-1)} \frac{\Gamma\left(\frac{1}{2}+\frac{1}{2 p}\right)}{\Gamma\left(\frac{1}{2 p}\right)} z_T^{1-p} \sim \frac{1}{\Gamma^{p-1}},
    \label{eq:Coulombbehavior}
\end{equation}
where we used \eqref{eq:proportional}.  The $\Gamma^{1-p}$ scaling is Coulomb law. For the special case of $p=1$, the first term in the series expansion for \eqref{eq:Coulombbehavior} cancels with the first term in \eqref{eq:actionintegral}, and the leading contribution goes like $\log(\Gamma)$. 

\subsection{The Critical Angle for the RS/KR Quark Potential}

Here we introduce a KR brane and repeat the computation for the defect-boundary potential, which we will loosely continue to refer to as the quark-antiquark potential.
The brane angle $\theta$ is an order parameter for the Wilson surface associated with the  $p-$form symmetry; we will demonstrate a transition from perimeter to Coulomb behavior at exactly $\theta_{c,p}$.

\begin{figure}
    \centering
\includegraphics[width=\linewidth]{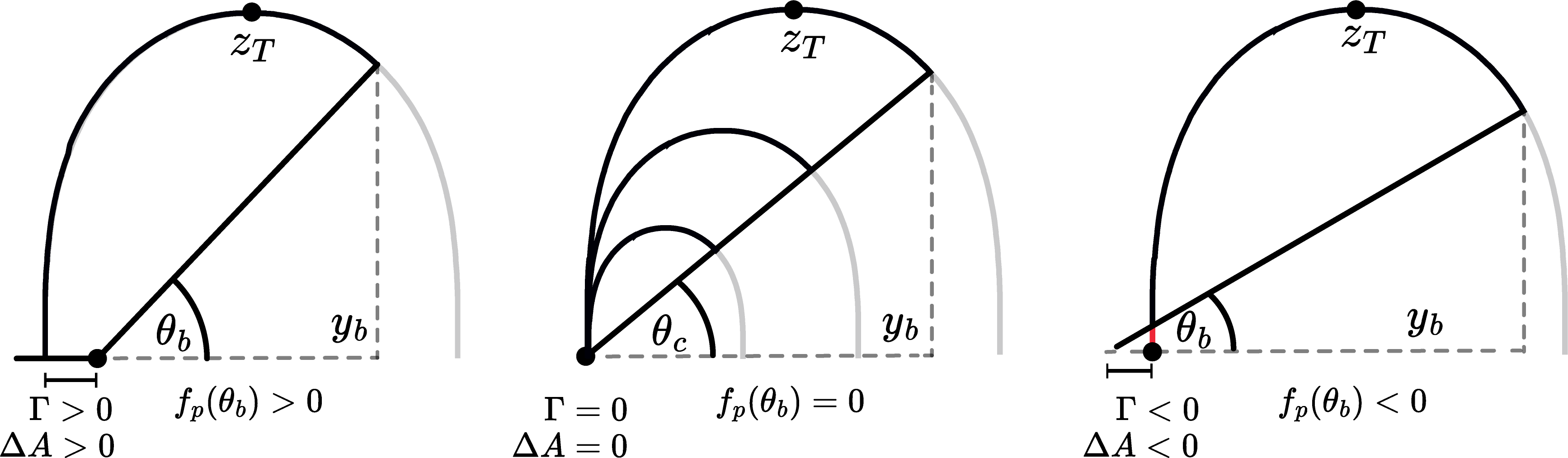}
    \caption{The physical interpretation of $f_p(\theta)$ is that it determines whether the extremal surface undershoots $(f_p(\theta)<0)$ or overshoots $(f_p(\theta)>0)$ the defect at $\Gamma=0$. On the LHS, above the critical angle, the surface overshoots the defect when $f_p(\theta)>0$, which gives $\Gamma>0$ and $A_p<0$. In the middle, at the critical angle $\theta_{c,p}$, $f_p(\theta)= A_p = \Gamma = 0$. Note the scale invariance at the critical angle. On the RHS we are below the critical angle; we have $f_p(\theta)<0$ and $\Gamma<0$, so the brane-anchored surfaces cease to exist.}
    \label{fig:shooting}
\end{figure}

As shown in Figure \ref{fig:Wilson_Line}, we must integrate from a point on the brane $(y_b, z_b)$, to the turnaround point at $z=1$, and down the other branch to $z=0$. Note $y_b$ is negative:
\beq \Gamma = z_T \underbrace{\left ( 2 \int_0^1 dz\, \frac{z^p}{\sqrt{1-z^{2p}}} - \int_{0}^{z_b} dz \, \frac{z^p}{\sqrt{1-z^{2p}}}  
+ y_b \right )}_{f_p(\theta)}.
\label{wz0relation}
\eeq
As illustrated in Figure \ref{fig:shooting}, $f_p(\theta)$ determines whether the extremal surface undershoots $(f_p(\theta)<0)$ or overshoots $(f_p(\theta)>0)$ the point at $z=0$. If the surface undershoots the point at $z=0$, no surface ending on the brane exists for this given value of $\Gamma$. This suggests the potential will vanish when $f_p(\theta)=0$ -- we will find that it does, and we will uncover interesting behavior at this transition point. 

The regularized area, $\mathcal{A}_p(\theta)$, vanishes for $\Gamma=0$ if the standard orthogonal boundary condition -- a consequence of ensuring the action is minimized even when varying the endpoint on the brane \cite{Chen:2020uac}  -- is obeyed at $z_b$. Branes are embedded along radial lines in Poincare coordinates \cite{Karch:2000ct}:
\begin{equation}
    \frac{y_b}{z_b} = \cot \theta.
\label{eq:orthog}
\end{equation}
The trajectory of the surface $y(z)$ crosses the radial line, between $z=0$ and $z_b$, at a right angle when $y'(z_b) = - \tan \theta $, so $z_b^p = \sin \theta$. Now we show this condition follows from demanding $\Gamma=0$, which implies $f_p(\theta)=0$:
\begin{equation}
y_b+2 \int_0^1 y^{\prime}(z) d z-\int_0^{z_b} y^{\prime}(z) d z=0.
\end{equation}
Defining $\alpha = \left(2 \sqrt{\pi}\right) \frac{ \Gamma\left(\frac{1+p}{2 p}\right)}{\Gamma\left(\frac{1}{2 p}\right)}$ for convenience, we find:
\begin{equation}
y_b= \alpha -\frac{z_b^{1+p}{ }_2 F_1\left(\frac{1}{2}, \frac{1+p}{2 p}, \frac{3+\frac{1}{p}}{2}, z_b^{2 p}\right)}{1+p}.
\label{eq:defectcon}
\end{equation}
Similarly, the condition for vanishing regulated area is $A_p(\theta) = 2 \int_0^1 \mathcal{A} \ d z-\int_0^{z_b} \mathcal{A} \ d z=0$. This gives: 

\begin{equation}
\alpha -z_b^{1-p}{ }_2 F_1\left(\frac{1}{2}, \frac{1+p}{2 p}-1, \frac{3+\frac{1}{p}}{2}-1, z_b^{2 p}\right)=0. 
\label{eq:areacon}
\end{equation}
Combining Equations \eqref{eq:defectcon} and \eqref{eq:areacon} yields a relationship between the coordinates at the crossing point:
\begin{equation}
\begin{split}
\frac{y_b}{z_b} &= z_b^{-p} {}_2F_1\left(\frac{1}{2}, \frac{1+p}{2 p}-1, \frac{3 p+1}{2 p}-1, z_b^{2 p}\right)\\ & \ \ -\frac{z_b^{p}{ }_2 F_1\left(\frac{1}{2}, \frac{1+p}{2 p}, \frac{3 p+1}{2 p}, z_b^{2 p}\right)}{1+p}.
\label{eq:messycon}
\end{split}
\end{equation}
A trick is to take the derivative with respect to $z_b$ and then integrate. The additive constant $c$ can be eliminated by noting $y_b(1)=c$ and computing \eqref{eq:messycon} at $z_b=1$:

\begin{equation}
    y_b(z_b)=\frac{\sqrt{1-z_b^{2 p}}}{z_b^{p-1}}.
    \label{eq:branesol}
\end{equation}
The physical interpretation is that $y_b$ vanishes when the turnaround point lies on the brane; i.e., the solution has fallen into the defect. 
The orthogonality condition, which we wanted to show, then follows from  \eqref{eq:orthog} and \eqref{eq:branesol}:
\begin{equation}
    \cos\theta_b = \sqrt{1 - z_b^{2p}} \implies z_b^p = \sin \theta_b.
\end{equation}
The \textit{critical angle} $\theta_{c}(p) \equiv \theta_{c,p}$ satisfying this condition remains interesting in a broader context. Using \eqref{eq:orthog}, \eqref{eq:messycon}:  
\begin{equation}
\begin{split}
    \cot \theta_c - &\frac{2 \sqrt{\pi } \Gamma \left(\frac{p+1}{2 p}\right) }{\Gamma \left(\frac{1}{2 p}\right) \sin ^{\frac{1}{p}}(\theta_c )} = \\&\frac{\sin (\theta_c ) \, _2F_1\left(\frac{1}{2},\frac{p+1}{2 p};\frac{1}{2} \left(3+\frac{1}{p}\right);\sin ^2(\theta_c )\right)}{p+1}.
    \label{eq:critical_angle}
\end{split}
\end{equation}
From \eqref{eq:critical_angle} the critical angle $\theta_{c,p}$ can be found numerically \cite{Geng:2020fxl}.
It vanishes for $p=1$ and approaches  $\pi/2$ monotonically as $p \rightarrow \infty$. Its value is unchanged at finite temperature.

Now we compute the regulated action. For $p>1$:
\begin{equation}
\begin{split}
A_p(\Gamma) &= 
z_T^{1-p} \left ( \int_{z_b}^1dz \,  {\cal A} + \int_{\epsilon}^1 dz \, {\cal A}
 - \frac{\epsilon^{1-p}}{p-1} \right ) \\
 &= \left(\frac{\Gamma}{f_p(\theta)}\right)^{1-p} \left ( \int_{z_b}^1dz \,  {\cal A} + \int_{\epsilon}^1 dz \, {\cal A}
 - \frac{\epsilon^{1-p}}{p-1} \right ),
 \label{eq:Coulomblaw}
\end{split}
\end{equation}
where $\epsilon = \tilde{\epsilon}/z_T$ is the rescaled cutoff. By using \eqref{wz0relation} we have tacitly assumed vanishing temperature. The renormalized $A_p(\Gamma)$ is finite.

For $\Gamma \neq 0$ the regulated area once again scales as $\frac{1}{\Gamma^{p-1}}$. However, Coulomb law goes unrealized below $\theta_{c,p}$ because $f_p(\theta)<0$; the surfaces connecting to the brane cease to exist. This is perimeter law because the potential between quark/defect and boundary does not depend on the distance to the boundary when the brane lies below the critical angle $\theta_{c,p}$. It is remarkable that the potential vanishes precisely when $f_p(\theta)$ changes sign.

Above $\theta_{c,p}$ the surfaces connecting to the brane have 
$A_p<0$, which dominates the disconnected configuration. Thus Coulomb law is realized in this regime. This follows because the area difference vanishes at the critical angle; increasing the brane angle reduces the regulated area.

This transition from Coulomb law above $\theta_{c,p}$ to perimeter law below $\theta_{c,p}$, with the brane angle acting as an order parameter, is one of the primary results of this letter. Many aspects of this transition are universal, as they apply to the near-boundary region in every asymptotically AdS geometry. As mentioned in the introduction, this transition applies equally well to the potential between two quarks screened by a brane in an ICFT.

The only option from a field-theory point of view is Coulomb law because the boundary preserves scale invariance. The transition we have uncovered simply indicates that the coefficient of the Coulomb term vanishes when the brane lies below $\theta_{c,p}$. This transition was not anticipated from field theory considerations alone.

\section{Screening at Finite Temperature}

Here we show that Coulomb law is not realized below the critical angle at finite temperature. The simplest way to extend our results to theories at finite temperature is to place the field theory on a non-dynamical black hole background, where a position-dependent finite temperature emerges because the field theory Hawking radiates. 

For simplicity we use the ``planar black string" metric for our bulk geometry, where $u_h$ is the horizon radius:

$$
ds^2 = \frac{1}{u^2 \sin^2\mu}\left[-h(u)dt^2 + \frac{du^2}{h(u)} + u^2 d\mu^2 + d\vec{x}^2\right]
    \label{bsMetL1},
$$
where $h(u) = 1 - \frac{u^{d-1}}{u_{\text{h}}^{d-1}}$ and $u \equiv \frac{1}{r}$ is measured from the boundary of AdS. Recall that for the zero temperature case, there was degeneracy between different $d$ for the same $p$; this is is clearly broken at finite temperature.

\begin{figure}
    \centering
    \includegraphics[width=\linewidth]{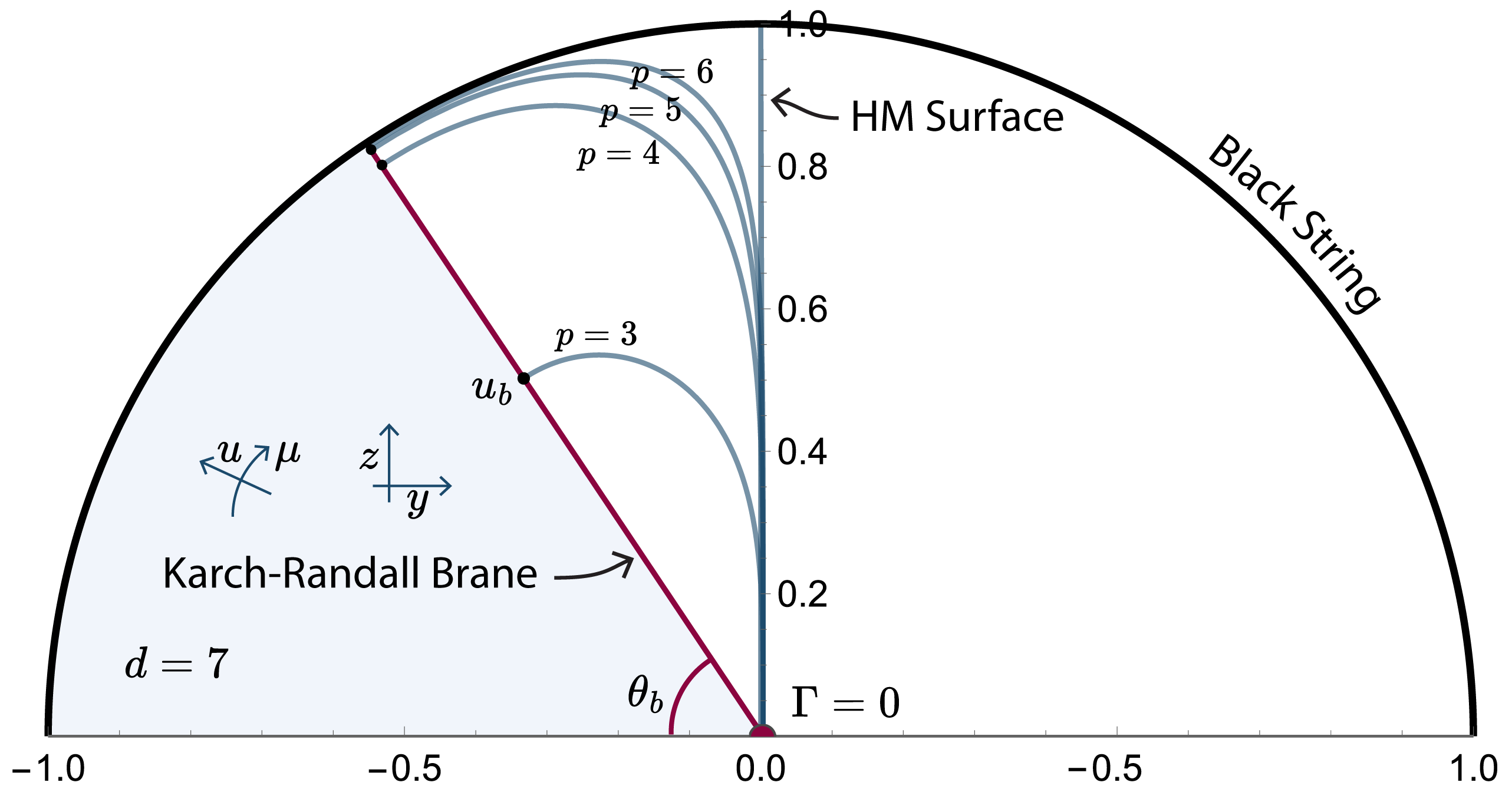}
    \caption{Here we illustrate the behavior at the critical angle $\theta_{c,p}$ at finite temperature for $p$-dimensional extremal surfaces dual to $p-1$ dimensional defect operators. The background geometry is the $d=7$ dimensional black string. Spatial Wilson surfaces anchored to the defect ($\Gamma=0$) touch the brane at the critical anchor $u_b$. This brane angle is slightly below the critical angle for $p=3$ surfaces: $\theta_{c,3} \approx .98687$, and above $\theta_{c,1}$ and $\theta_{c,2}$, which means the $p=1$ and $p=2$ dimensional surfaces have vanishing critical anchors. Only their area differences diverge with (generalized) Coulomb law scaling. }
\label{fig:blackstringwilsonlines}
\end{figure}

In the field theory we are still looking at time independent configurations parallel to the defect; in the $u$-$\mu$ coordinates we want $u=u(\mu)$ with $u(0)=\Gamma$. In this finite temperature case the degeneracy between spacelike and timelike surfaces is \textit{also} broken. For a spacelike surface we take, as before, $t=0$ and $x_i=0$. Then:
\begin{align}
A
&= V \int \frac{d\mu}{[u \sin\mu]^{p}} \sqrt{\frac{(u')^2}{h[u]} + u^2}.
\end{align}
For temporal Wilson surfaces, the timelike surface treats one more of the $x_i$ as a transverse coordinate, which simply introduces an extra factor of $\sqrt{h(u)}$ into the action. 

Figure~\ref{fig:blackstringwilsonlines} shows extremal surfaces that may end on the KR brane at a critical anchor $u_b$. Coulomb scaling occurs only when the brane satisfies $\theta_b > \theta_{c,p}$; the resulting surfaces shrink to infinitesimal size as $\Gamma \rightarrow 0$. Otherwise the saddle attaches at finite $u_b$ with $u_b \rightarrow 0$ as $\theta \rightarrow \theta_{c,p}$. Figures, boundary conditions, and additional details can be found in Appendix \ref{numericalverification}.

\subsection{The Area Difference at the Critical Angle}

For the black string the HM surface is a non-trivial equal partner with the surface connecting to the brane. The area of the HM surface anchored to the defect is computed for generic $p$ and $d$ at finite temperature. This suffices to determine the area difference at the critical angle for $\Gamma=0$ because the regulated area of a surface shrinking into the defect vanishes, i.e. $\Delta A_c = - A_{HM}$. 

This calculation is for a spatial Wilson line but the computation for the temporal Wilson line gives the same regulated area; see Appendix \ref{temporal}. The area functional can be straightforwardly written using $\mu(u)$ as:

\begin{equation}
    A =\int \frac{d u}{(u \sin \mu)^{p}} \sqrt{u^2 \mu^{\prime}(u)^2+\frac{1}{h(u)}}.
\end{equation}
The HM surface plunges directly into the black hole when $\Gamma=0$, so $\mu'(u)=0$. Since $\mu = \pi/2$ along its trajectory, the area of the HM surface is:
\begin{equation}
    A_{HM} = \int_0^1 \frac{d z}{z^{p}} \left(\sqrt{\frac{1}{h(z)}}\right).
    \label{eq:areaHM}
\end{equation}
For the planar AdS black hole we have $h(z) = 1 - z^{d-1}$. 

We regulate this formally divergent integral by a change of variables adapted to the black-hole shadow. For an observer at a distance $r_0$ from the photon ring at $r_{\text{ps}}$, the angular extent $\alpha$ of the black hole shadow is $\sin (\alpha)=\sqrt{\frac{h\left(r_0\right)}{h\left(r_{\mathrm{ps}}\right)}}$. For the photon ring in our setup, an observer located on the slice along the HM surface is a distance $z$ from the photon ring at the AdS boundary:
\begin{equation}
    \csc(\alpha) = \sqrt{\frac{1}{h(z)}}.
    \label{eq:coordinatesub}
\end{equation}
This coordinate transformation applied to \eqref{eq:areaHM} gives $A_{HM} = \frac{2}{d-1} \int_0^{\pi/2} \cos(\alpha)^n d\alpha$,
where $n=2 \left(\frac{1-p}{d-1}\right)-1$. For analytical continuation in $n$, the integral is simply
$
    \int_0^{\pi / 2} \cos (\alpha)^n d \alpha = \frac{\sqrt{\pi} \Gamma\left(\frac{n+1}{2}\right)}{2 \Gamma\left(\frac{n}{2}+1\right)}
$. Then $\Delta A$ at $\theta_{c,p}$ is:
\begin{equation}
     \Delta A_c = - A_{HM} = \frac{\sqrt{\pi}}{1-d} \frac{\Gamma\left(\frac{1-p}{d-1}\right)}{\Gamma\left(\frac{1-p}{d-1}+\frac{1}{2}\right)}.
    \label{eq:areadiffanswer}
\end{equation}
This is infinite when $\frac{p-1}{d-1}$ is an integer and vanishes when $\frac{p-1}{d-1}$ is a half-integer.  Since $p<d$ for all solutions of interest  this integral is finite. This expression is verified numerically in Appendix \ref{numericalverification}, see Figure \ref{fig:Coulomblaw}. The limiting case $p \rightarrow 1$, where the critical angle vanishes, needs the counter-term $\frac{1}{p-1}$. Taking the limit, some remarkable simplification gives $\Delta A_{c,p=1} =  \frac{ \log(4)}{1-d}$. 

\section{Discussion}

In this letter we used Karch-Randall (KR) branes to screen the potential between quarks and extended this to include $p-1$ dimensional defect operators. This generalizes a key result showing that Wilson lines and surfaces exhibit Coulomb scaling laws in AdS/CFT at zero temperature \cite{Maldacena:1998im,Rey:1998ik}. In this simple setup, the ``island" phase transition that appears in standard discussions of entanglement islands emerges for Wilson and 't Hooft operators. The potential between the quarks vanishes at a critical angle for the brane $\theta_{c,p}$; the Wilson line connecting the quarks detaches from the branes and an additional phase transition occurs between Coulomb law and perimeter law, with the brane angle acting as an order parameter. This same effect occurs for more general $p-1$ dimensional defect operators, both in ICFTs and BCFTs. 

This phenomenon was also explored at finite temperature, where we introduced a regularization technique inspired by the photon ring to obtain closed-form expressions for divergent integrals using analytic continuation. This was used to compute the action at finite temperature at $\theta_{c,p}$ for $p-1$ dimensional defect operators. This simple procedure could prove useful in other higher-dimensional setups, and may indicate a deeper physical principle that has not yet been uncovered. Similarly at finite temperature,  we find Coulomb scaling laws go unrealized below the critical angle $\theta_{c,p}$ for the brane. 

\section{Acknowledgments}
Special thanks to Jacques Distler and Amir Raz for critical questions and useful discussions. MR also thanks Hao-Yu Sun and Zixia Wei for helpful comments.
This work was supported, in part, by the U.S.~Department of Energy under Grant DE-SC0022021 and by a grant from the Simons Foundation (Grant 651678, AK). MR was also supported by a Dissertation Fellowship from CNS, an OGS Summer Fellowship, the Dean's Strategic Fellowship, and the GRASP initiative at Harvard University.

\clearpage

\appendix

\section{Temporal Wilson Surfaces}
\label{temporal}
In the main text we gave an analytic expression for the area difference between connected and disconnected minimal surface in the case of a spatial Wilson surface.
Here we give the analogous argument for the temporal Wilson surface. The action in Eq. \eqref{eq:action} picks up an extra factor of $\sqrt{h(u)}$:
\begin{equation}
    S= V \int \frac{d\mu}{[u \sin\mu]^{p}} \,
\sqrt{h[u]} \sqrt{\frac{(u')^2}{h[u]} + u^2}.
\end{equation}
Rewriting this in terms of $\mu(u)$, minor variations on the same argument give for $\Delta A$:
\begin{equation}
    \Delta A = - \int_0^1 \frac{d z}{z^{p}} \left(\sqrt{h(z)}\right).    \label{eq:areadiffregtemporal}
\end{equation}
Notice in this case there is simply an extra factor of $\sqrt{h(z)}$. In fact, this is $h(r) = \frac{f(r)}{r^2}$, but in coordinates $h(u)$ where $u= \frac{1}{r}$. The photon ring is located at $u_{ps}$ where $h'(u_{ps})=0$. These coordinates automatically regulate the integral: 
\begin{equation}
\begin{aligned}
    A_{HM} &= \int_{0}^{\pi/2} \frac{2\left(\sin (\alpha)^2\right)^{\frac{1-p}{d-1}} \csc (\alpha)}{d-1} d\alpha \\ &= \frac{2}{d-1} \int_0^{\pi/2} \sin(\alpha)^{n} d\alpha.
\end{aligned}
\end{equation}
Once again we have $n=2 \left(\frac{1-p}{d-1}\right)-1$. For analytical continuation in ${n}$, we have $\int_0^{\pi / 2} \sin (\alpha)^{n} d \alpha = \frac{\sqrt{\pi} \Gamma\left(\frac{{n}+1}{2}\right)}{2 \Gamma\left(\frac{{n}}{2}+1\right)}$. Despite the differences in this calculation the area difference is the same as for the spatial Wilson surface:
\begin{equation}
    \Delta A_c = \frac{\sqrt{\pi}}{1-d} \frac{\Gamma\left(\frac{1-p}{d-1}\right)}{\Gamma\left(\frac{1-p}{d-1}+\frac{1}{2}\right)}.
    \label{eq:TWLarea}
\end{equation}
The following may shed some light on why finite answers can be obtained straightforwardly using this method. Instead of implementing a non-covariant UV cutoff, the calculation is rephrased covariantly in terms of the shadow of the planar black hole. It would be interesting to understand this procedure in greater detail.

\section{Numerical Verification}
\label{numericalverification}
Here we explain how the results described in the main text were numerically verified. 
\subsection{Boundary Conditions}

To obtain the extremal surfaces, we must first vary the action to obtain the equations of motion; specifics concerning the detailed equations that follow can be found in \cite{Geng:2020fxl,Geng:2021hlu}. The Neumann boundary condition for surfaces ending on the brane is the same as in \cite{Geng:2020fxl,Geng:2021hlu}. The Neumann boundary condition for surfaces ending on the horizon is needed at finite temperature. The surface must cross the horizon smoothly in local coordinates; that is, $\mu'(u) = 0$ and $\mu''(u) = 0$ at the horizon. 

Here is how the boundary condition is found for the temporal Wilson surfaces at finite temperature. The Euler-Lagrange equations are first solved for $\mu''(u)$.  Expanding that expression in powers of $u$ near the horizon gives: 

\begin{equation}
    \mu''(u) \approx \frac{(u_h-d u_h) \mu '(u_h)+p \cot (\mu (u_h))}{(d-1) u_h (u-u_h)}.
\end{equation}
Demanding $\mu''(u) = \mu'(u) = 0$ at the horizon gives the boundary condition:
\begin{equation}
    \mu '(u_h) =  \frac{p \cot (\mu (u_h))}{(d-1) u_h}.
\end{equation}
This procedure applied to the spatial Wilson surfaces gives:
\begin{equation}
    \mu '(u_h) =  \frac{p \cot (\mu (u_h))}{2(d-1) u_h}.
\end{equation}

\subsection{Critical Anchors and Phase Transitions}

\begin{figure}
    \centering
    \includegraphics[width=\linewidth]{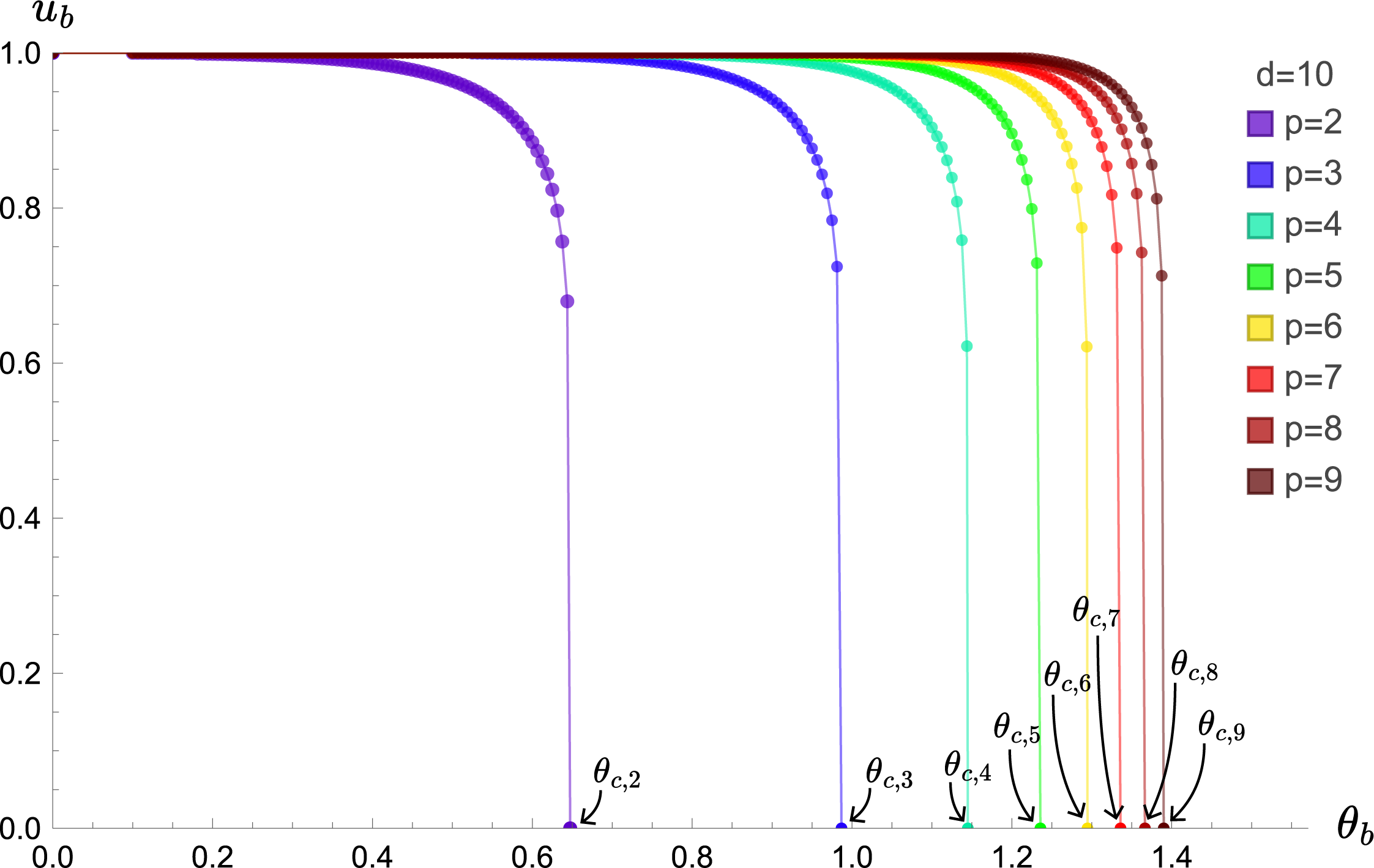}
    \caption{Critical anchors $u_b$ are shown for the ($d=10$) dimensional black string geometry as a function of brane angle $\theta_b$ and surface dimension $p$. The $u_b$ vanish above the critical angle $\theta_{c,p}$. The same limiting behavior at $\theta_{c,p}$ was found for $p-$dimensional surfaces, with $p<d$, for each bulk dimension $d>2$. Similar comments apply to Figures \ref{fig:constantd} and \ref{fig:constantp}; $u_b$ is monotonically increasing in both $p$ and $d$, and monotonically decreasing in $\theta_b$. This phenomenon occurs in every asymptotically AdS geometry with an embedded Karch-Randall brane. The surface for $p=1$ is not pictured because $\theta_{c,1}=0$. }
    \label{fig:constantd}
\end{figure}

\begin{figure}
    \centering
    \includegraphics[width=\linewidth]{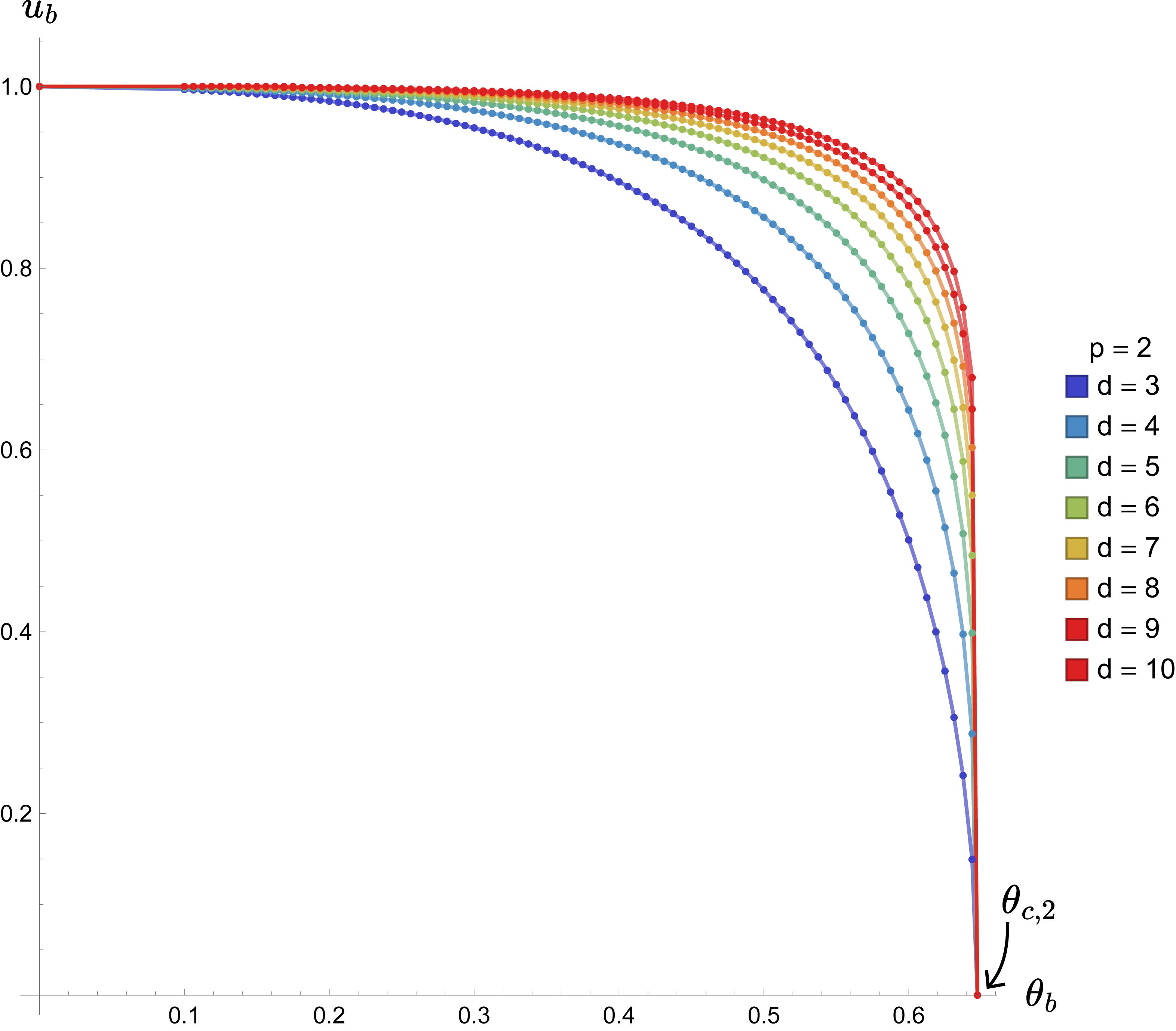}
    \caption{The critical anchors $u_b$ are displayed as a function of brane angle $\theta_b$ and bulk dimension $d$ for surface dimension $p=2$ in the black string geometry. The $u_b$ vanish at the critical angle $\theta_{c,2}$. The same limiting behavior was found at $\theta_{c,p}$ for $p$-dimensional extremal surfaces, with $p<d$, for each bulk dimension $d$.}
    \label{fig:constantp}
\end{figure}

Here it is convenient to use the shooting method, where the endpoint on the brane $u_b$ is varied to ``shoot" $u(\pi)=\Gamma$ while enforcing the Neumann boundary condition on the brane $u'(\theta_b)=0$. This procedure, which is roughly analogous to hitting a target off the coast at position $\Gamma$ with a projectile, can be straightforwardly implemented using the shooting method of NDSolve in Mathematica. 

This method applied to spatial Wilson surfaces at $\Gamma=0$ yields Figure \ref{fig:constantd} and Figure \ref{fig:constantp}, where the critical anchors $u_b$ behave as we have predicted in the main text. The critical anchor $u_b$ for a $p$ dimensional extremal surface vanishes above the critical angle $\theta_{c,p}$. For spatial Wilson surfaces the value of $u_b$ is monotonically increasing in $p$ and $d$ and monotonically decreasing in $\theta_b$. Spatial Wilson surfaces continue to exist below the critical angle at finite temperature. This is unsurprising for spatial Wilson surfaces because the black hole horizon at $u_b = u_h = 1$ is an extremal surface. 

Interestingly, the temporal Wilson surfaces do not exist below the critical angle on the black string background. Then the transition from Coulomb law to perimeter law that occurs for the potential at zero temperature also arises in this case, where the black hole on the brane is planar. However, we have found these surfaces \textit{continue} to exist for the \textit{global} black string below the critical angle; in that case \cite{Karch:2023ekf} the black hole on the brane is a sphere. Such explorations are left for future work.

\subsection{Computing the Area Differences}
\label{app:Area}

Here we show how the area differences between brane-anchored and horizon-plunging (HM) surfaces at finite boundary anchor $\Gamma$ are determined. First the shooting method is used to obtain the solutions for each $\Gamma$ and the area difference is computed numerically. Numerical integration is less reliable near the AdS boundary which is a singular point of the Euler-Lagrange equations. The Graham-Witten procedure \cite{Graham:1999pm} is used to compute the area difference in that regime. 

This computation in higher dimensions has interesting structure that will be described more thoroughly in an upcoming note.  The Euler-Lagrange equations are used with the following ansatz for $\mu(u)$: 

\begin{figure}
    \centering
    \includegraphics[width=\linewidth]{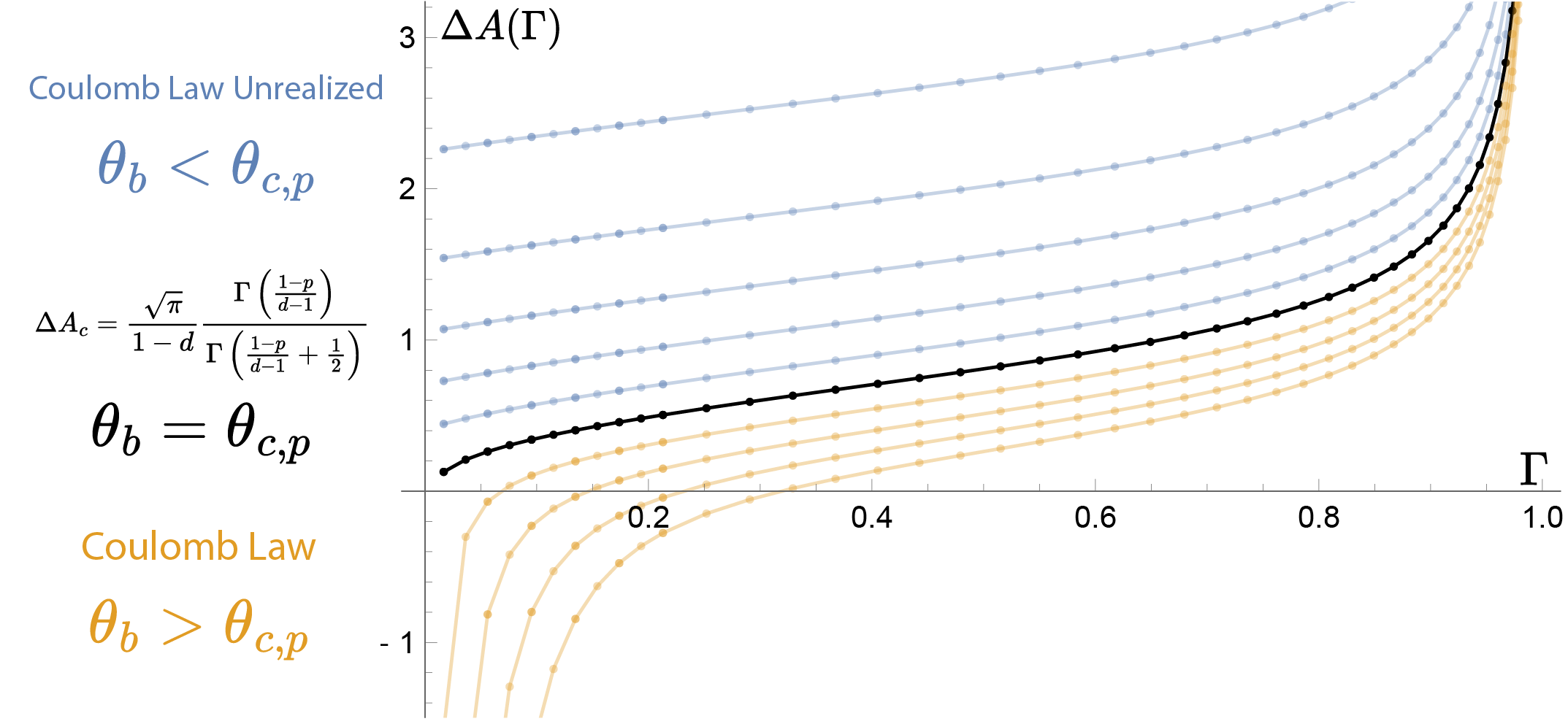}
    \caption{The area differences between brane-anchored and horizon-plunging (HM) surfaces at finite boundary anchor $\Gamma$ were computed at finite temperature. Here the numerical results for $p=3$, $d=5$ are shown. The area difference at the critical angle in the limit where $\Gamma \rightarrow 0$ is in agreement with Eq. \eqref{eq:areadiffanswer}. Coulomb scaling is realized only above the critical angle $\theta_{c,p}$ for the brane. The area difference below $\theta_{c,p}$ is finite; the surfaces touch the brane at a finite critical anchor $u_b$.}
    \label{fig:Coulomblaw}
\end{figure}

\begin{equation}
    {c_0} + c_{40} (\pi -\mu )^4+c_{20} (\pi -\mu )^2+c_{41} (\pi -\mu )^4 \log (\pi -\mu ).
\end{equation}
The objective is to determine the coefficients near the singular point at the AdS boundary $\mu=\pi$. Expanding the Euler-Lagrange equations around this point gives $c_{20}$: 
\begin{equation}
c_{20} =\frac{1}{2} u_h^{-d} \left(c_0^3 u_h^d+c_0 u_h^d-u_h c_0^d\right).
\end{equation}
Inserting $c_{20}$ and then expanding the Euler-Lagrange equations to higher order gives a complicated expression for $c_{41}$ in terms of $c_{40}$. This suffices to find an expression for $\mu(u)$ and $\mu'(u)$ in the near-boundary region in terms of these coefficients. Some cutoff region for the numerics near the AdS boundary is needed, such as $\mu = \pi - .001$. The next step is to determine $u(\mu)$ and $u'(\mu)$ numerically at the cutoff and solve algebraically for $c_{0}$ and $c_{40}$.

Once the coefficients $c_{i j}$ are determined the approximate area functional is integrated. The resulting expression is used to compute the area difference between the surfaces in the cutoff region. To obtain the full area difference, this is combined with the area differences computed from the numerical solutions up to the cutoff. 

The area differences at the critical angle for the finite temperature case in the limit where $\Gamma \rightarrow 0$ are in agreement with our predictions. Coulomb scaling is realized only above the critical angle $\theta_{c,p}$ for the brane, where the $p$-dimensional surfaces can shrink to infinitesimal size. The area difference below $\theta_{c,p}$ is finite because the surfaces touch the brane at a finite critical anchor $u_b$. 

\section{Type IIB Realization}
\label{TypeIIBRealization}

While the geometries we discussed involving KR branes are ``bottom-up", that is solutions to Einstein equations  with idealized matter sources without known UV completion or known holographic dual, they do capture qualitative aspects of actual ``top-down" solutions in type IIB supergravity, the low energy limit of type IIB string theory. The full IIB geometry is much more complicated, but analytic solutions are known
\cite{DHoker:2007hhe,DHoker:2007zhm,Aharony:2011yc,Assel:2011xz}. 
These are 10d solutions with a non-compact 5d part. The dual field theories are well understood and had prior been analyzed in great detail by Gaiotto and Witten \cite{Gaiotto:2008sa} as we will describe in more detail below. Instead of a bulk spacetime ending on KR brane, the geometry truncates because some cycle on the internal space collapses to zero size. The resulting geometries are sometimes referred to as bagpipes \cite{Bachas:2018zmb}, as a non-compact AdS$_5$ $\times$ $S^5$ throat pulls out of a compact AdS$_4$ $\times_w$ ${\cal M}^6$ ``bag" that contains 5-brane sources. There is no limit in which the top-down models reduce to the bottom-up models, but they do exhibit qualitatively very similar behavior, in particular they exhibit the critical angle phenomenon.

The critical angle for the case of an RT surfaces was first investigated from the top-down in this geometry by Uhlemann \cite{Uhlemann:2021nhu} and later in
\cite{Demulder:2022aij}.
The analog of the critical angle in these top-down constructions is best understood from the point of view of the dual field theory. One describes the theory living on D3 branes ending on a stack of D5 and NS5 branes, that is ${\cal N}=4$ super Yang Mills living on half space, coupled to a 3d supersymmetric CFT with a quiver gauge theory whose degrees of freedom (as measured by the sphere free energy) grow with the number of 5-branes involved. The rank of the 4d gauge group sets the curvature radius of the asymptotic AdS$_5$ spacetime, whereas the free energy of the 3d quiver gauge theory determines the curvature radius of the AdS$_4$ slice at the end of space. Since the former is controlled by the number of 3-branes, and the latter is controlled by the number of 5-branes, they can be independently dialed, and the analog of the angle is the ratio of 3d to 4d degrees of freedom. A very large number of 3d degrees of freedom corresponds to a very large AdS$_4$ curvature radius and so corresponds to the small angles of the bottom-up model. It is this 3d dominated limit that yields a small graviton mass. Reassuringly, the studies of \cite{Uhlemann:2021nhu,Demulder:2022aij} found that, at least for RT surfaces, varying this ratio in these top-down geometries leads to a transition with exactly the same properties of the critical angle transition described in this work.
It would be very interesting to study probes realizing defects with different values for $p$ in these top-down constructions.

\bibliographystyle{apsrev4-1}
\bibliography{references.bib}

\end{document}